\documentclass[prd, nofootinbib,aps,superscriptaddress,tightenlines,preprintnumbers]{revtex4}
\usepackage{latexsym}
\usepackage{amsmath}
\usepackage{amssymb}
\usepackage{amsfonts}
\usepackage{comment}
\usepackage{graphicx}
\usepackage{verbatim}
\usepackage{subfig}
\usepackage[countmax]{subfloat}

\newcommand{\be}{\begin{equation}}
\newcommand{\ee}{\end{equation}}
\newcommand{\bea}{\begin{eqnarray}}
\newcommand{\eea}{\end{eqnarray}}

\begin{document}
\title{Gravitational Field Inside and Near a Caustic Ring}

\author{Heywood Tam}
\affiliation{Department of Physics, University of Florida, Gainesville, Florida 32611, USA }

\begin{abstract}
We present an analytic calculation of the gravitational field inside and near a caustic ring of dark matter.  The calculation may facilitate N-body simulation studies on the effects that dark matter caustics have on galaxy formation.  
\end{abstract}

\date\today
\maketitle

\section{Introduction}

One mystery in modern cosmology is the identity of the dark matter, which constitutes 23\% of the energy density of the current universe.  By now, a host of observations as well as numerical simulations of structure formation have established that dark matter must be cold and collisionless nonbaryonic particles that interact primarily via the gravitational interaction.   Since no known particles have these properties, one has to turn to physics beyond the Standard Model for suitable candidates.   Currently, the leading candidates of cold dark matter (CDM) are axions and weakly-interacting massive particles (WIMPs).  The former originates from Peccei and Quinn's dynamical solution to the strong CP Problem \cite{Peccei:1977hh,Peccei:1977ur,Weinberg:1977ma,Wilczek:1977pj}, while the latter naturally arises from R-parity-conserving supersymmetry. 

For a long time, it is believed that the two candidates are indistinguishable on observational grounds, even though fundamentally they are very different.  The reason for this has to do with the fact they can both be viewed as classical approximations of quantum fields, albeit in different limits.   WIMPs, as fermions, are in the regime of classical particles, in which we take $\hbar \rightarrow 0$ and $\omega, k \rightarrow \infty$ while keeping their energy $E=\hbar \omega$ and momentum $p = \hbar k$ finite.  On the other hand, axions, as bosons, reside in the classical field regime, in which we take $\hbar \rightarrow 0$ and the occupation number $\mathcal{N}\rightarrow\infty$ for constant $\omega$ and $k$ to keep $E=\mathcal{N}\hbar\omega$ and $p = \mathcal{N} \hbar k$ fixed \cite{Erken:2011dz}.  Hence, both behave as a pressureless perfect fluid, and have similar properties as far as large scale structure is concerned.  For example, it has been demonstrated that the evolution of cosmological perturbations in an axion BEC and ordinary cold dark matter are practically indistinguishable on scales of observational interest provided the BEC does not rethermalize \cite{Bianchi, Sikivie:2009qn, Hwang:2009js}.  As such, it is widely held that the verdict on the true identity of dark matter can come only with a direct detection in experiments (there is of course the possibility that neither candidate is correct).  Currently, a large number of experiments worldwide are actively searching for both candidates, so far with negative results.

Recently, however, it has been pointed out that dark matter axions, being a Bose-Einstein condensate (BEC) \cite{Sikivie:2009qn}, could potentially give rise to novel avenues to differentiate them from WIMPs.  If, for example, thermal contact were established between the axions and photons after big-bang nucleosynthesis and before recombination, various cosmological parameters (namely the effective number of neutrinos $N_{eff}$ and the primordial abundance of light elements such as $^4$He, $^3$He, $^7$Li, etc.) would change \cite{Erken:2011vv}.  Hence, a more precise measurement of $N_{eff}$ by the {\it Planck} mission in the near future might help shed light on the identity of dark matter. 

In addition to cosmological parameters, the study of the catastrophe structure of inner caustics of galactic halos may provide yet further clues \cite{Natarajan:2005ut}.  Caustics are formed as galaxies accrete dark matter particles in their vicinity during structure formation.  They are surfaces of high density which can be viewed as the envelope of particle trajectories, as dark matter particles fall in and out of the gravitational well.  This process of particle infalling and outfalling occurs repeatedly, and a set of inner (as well as outer) caustics form.  It turns out that there exists a dichotomy in the classification of the inner caustics in terms of their catastrophe structure, depending on the distribution of angular momentum of the infalling particles.  The first case is characterized by ``net overall rotation" ($\nabla \times \vec v \ne 0$), and the corresponding inner caustic is a closed tube whose cross section is a section of the elliptic umbilic ($D_{-4}$) catastrophe, called a ``caustic ring''. The second case is characterized by irrotational flow ($\nabla \times \vec v = 0$), which gives rise instead to inner caustics with a tent-like structure.  Tidal torque theory predicts that ordinary cold dark matter such as WIMPs would be irrotational \cite{Natarajan:2005ut}, hence giving rise to tentlike caustics.  In contrast, ``net overall rotation'' ($\sim$ vortices) in an axionic BEC would result in ring caustics \cite{Erken:2011dz,Sikivie:2009qn}. 

For the first case (``net overall rotation"), the radii of the caustic rings $a_n$ can be predicted in terms of a single parameter, $j_{max}$, under the assumption that the time evolution of the galactic halo is self-similar \cite{Duffy:2008dk}. That is, the phase space structure of the halo is time-independent under appropriate rescaling of all distances, velocities, and densities \cite{Fillmore:1984wk, Bertschinger:1985pd, Sikivie:1995dp, Sikivie:1996nn, Duffy:2008dk}. If initially the overdensity in the halo has the profile $\delta M/M \propto M^{-\epsilon}$, where the parameter $\epsilon$ is estimated to lie in the range 0.25 to 0.35 from the observed power spectrum of density perturbations on galaxy scales, the prediction for $a_n$ is given by \cite{Sikivie:1997ng,Sikivie:1999jv,Duffy:2008dk}
\begin{equation}\label{radii}
a_n \simeq \frac{40 \mbox{kpc}}{n}\left(\frac{v_{rot}}{220 \mbox{km/s}}\right)\left(\frac{j_{max}}{0.18}\right),
\end{equation}
where $v_{rot}$ is the galactic rotation velocity and $n = 1,2,3,\ldots$

Intriguing hints for caustic rings with radii consistent with the prediction of Eq. \eqref{radii} have been found in a multitude of observations: the distribution of bumps in the rotation curve of the Milky Way (MW) \cite{Sikivie:2001fg}, and the appearance of a triangular feature in the IRAS map of the MW in the precise direction tangent to the nearest caustic ring \cite{Sikivie:2001fg}. Outside our home galaxy, further corroboration can be found in the statistical distribution of bumps in a set of 32 extended and well-measured galactic rotation curves \cite{Kinney:1999rk}; the recent rotation curve of the Andromeda galaxy, which shows three bumps at locations consistent with the prediction of \eqref{radii}; and gravitational lensing evidence in galaxy clusters \cite{Onemli:2007gm}.   Together, these observations lend credence to the argument that an axion BEC constitutes a substantial fraction of dark matter. 

Further evidence that may favor the existence of caustic rings is the recently-discovered Monoceros Ring of stars, an over-density of stars in the plane of the Galaxy at a galactocentric distance of $20$ kpc \cite{Newberg:2001sx, Yanny:2003zu}.  Their proximity to the $n=2$ inner caustic has led to speculations that their origin has to do with the presence of the ring \cite{Natarajan:2007xh}.  To test this claim, one will have to resort to numerical simulations involving the gravitational field near a caustic ring.  Unfortunately, the expression for the gravitation field is in general in the form of an integral, whose evaluation is computationally expensive, rendering numerical simulations difficult.  It is the goal of this paper to help circumvent this difficulty by evaluating the near-field caustic-ring gravitational field by analytical methods.  

This paper is structured as follows.  In Section II, we present our calculation of the gravitational field near a caustic ring (in the limit of infinite radius), specializing to the case where $\xi$, a parameter which we shall define and expect to be close to one, equals unity.  In Section III, we compare our analytic results with numerics, to establish the validity of our formulae in Section II.  

Finally, we note that, building on insights acquired in Section II, we can straightforwardly generalize our calculation to the case where $\xi \ne 1$. The computational method employed in Section II remains applicable, though the algebra becomes more cumbersome.  Since the generalization is not particularly illuminating, it is omitted in this paper. 

\section{Calculation}
\subsection{Theoretical Framework}
Much of the discussion in this subsection is taken from \cite {Natarajan:2007xh}.  Assuming axial symmetry and that the radius of the ring far exceeds the transverse dimensions of the caustic, the distribution of CDM near a caustic ring at some given time is given by
\begin{eqnarray} \label{rhoeqn}
\rho &=& a + \frac{1}{2}u(\tau-\tau_0)^2 - \frac{1}{2}s\alpha^2, \\ \label{zeqn}
z &=& b\alpha \tau,
\end{eqnarray}
where $\rho$ and $z$ are the usual cylindrical coordinates \cite{Sikivie:1999jv} (the angular coordinates $\phi$ is absent due to rotational symmetry).  The parameters $b, a, u, \tau_0, s$ characterize the caustic ring (see \cite{Natarajan:2007xh} for a good summary).  The variable $\tau$ is a temporal coordinate, and denotes the time when the CDM particle crosses the $z=0$ plane.  Hence $-\tau$ can be viewed as the age of the particle.  The variable $\alpha \equiv \frac{\pi}{2}-\theta$ is an angular coordinate, with $\theta$ being the polar angle of the particle at its last turnaround.  Hence we have two related sets of variables describing the dynamics of the particles: $(\rho, z)$ and $(\alpha, \tau)$. The location of the caustic is obtained from Eqs. \eqref{rhoeqn} and \eqref{zeqn} by solving for the values of $\alpha$ and $\tau$ at which the Jacobian determinant between the two sets of coordinates vanishes:
\begin{equation}
D(\alpha, \tau) = \mbox{det}\left(\frac{\partial (\rho, z)}{\partial (\alpha,\tau)} \right) = -b\left(u\tau(\tau-\tau_0) + s \alpha^2 \right) = 0.
\end{equation}
The density of the dark matter particles in physical space is given by \cite{Natarajan:2007xh}
\begin{equation} \label{density}
d(\rho, z) = \frac{1}{\rho} \sum_{j=1}^{N(\rho,z)} \frac{dM}{d\Omega d\tau}(\alpha,\tau) \left. \frac{\cos (\alpha)}{ |D(\alpha,\tau)|} \right|_{\alpha_j (\rho,z), \tau_j(\rho,z)},
\end{equation}
where $\alpha_j(\rho,z)$ and $\tau_j(\rho,z)$ are solutions to $\rho(\alpha,\tau)=\rho$ and $z(\alpha, \tau)=z$.   The number of solutions $N(\rho,z)$ represents the number of flows at location $(\rho, z)$.  $\frac{dM}{d\Omega d\tau}$ is the mass of dark matter particles falling in per unit solid angle and time. 

For convenience, we introduce two new variables
\begin{eqnarray} 
p &=& \frac{1}{2} u \tau_0^2 \\
q &=& \frac{\sqrt{27} b p}{4 \sqrt{us}}. 
\end{eqnarray}
The variables $p$ and $q$ have clear physical meanings: $p$ is the longitudinal dimension of the caustic (the caustic extends from $a$ to $a+p$ at $z=0$), while $q$ is the transverse dimension (the highest point of the caustic are at $z=q/2$).

Under the assumption that $p,q \ll a$ (i.e. infinite caustic radius), the gravitational field $\vec g$ is given by \cite{Natarajan:2007xh}
\begin{equation}
\vec g (\rho ,z) = -2 G \int d\rho' dz' d(\rho' , z') \frac{(\rho - \rho', z-z')}{(\rho - \rho')^2 + (z-z')^2},
\end{equation}
where $\rho'$ is integrated from $0$ to $\infty$ and $z'$ from $-\infty$ to $\infty$. 

Performing a coordinate transformation from $(\rho ,z)$ to $(\alpha, \tau)$ and using Eq. \eqref{density}, and ignoring the spatial variation of $\frac{dM}{d\Omega d\tau}$ over the size of the caustic, we have
\begin{equation} 
\vec g (\rho ,z) = -\frac{2G}{\rho}\frac{dM}{d\Omega d\tau} \int d\alpha d\tau \frac{(\rho - \rho (\alpha,\tau), z-z(\alpha,\tau) )}{(\rho - \rho(\alpha,\tau))^2 + (z-z(\alpha,\tau))^2}.
\end{equation}
To simplify the calculation, we introduce the following rescaled variables
\begin{eqnarray}
T &=& \frac{\tau}{\tau_0} \\
A &=& \frac{b \alpha}{u \tau_0} \\
Z &=& \frac{z}{p} \\
X &=& \frac{\rho - a}{p},
\end{eqnarray}
and the dimensionless parameter $\xi = su/b^2$.  The surface of the caustic is then given by:
\begin{eqnarray} \label{boundary}
X_s &=& (T-1)(2T-1) \\ \label{boundary2}
Z_s &=& \pm 2\sqrt{\frac{T^3(1-T)}{\xi}}
\end{eqnarray}

From Eqs. \eqref{boundary} and \eqref{boundary2}, we see that the caustic extends from $X=-1/8$ to $X=1$, independent of the value of $\xi$.  However, the highest (lowest) point of the caustic $Z_{max}$ ( $Z_{min}$) does depend on $\xi$, and is given by $\pm \frac{3}{8}\sqrt{\frac{3}{\xi}}$.  For $\xi =1$, the caustic possesses a $Z_3$ symmetry, as its three cusps rotate into one another under a rotation by $120 ^\circ$.  

The gravitational field of the caustic ring has the form
\begin{equation}
\vec g = -\frac{8\pi G}{\rho b}\frac{dM}{d\Omega d\tau}  (I_\rho(X,Z), I_z(X,Z)),
\end{equation}
where $I_\rho(X,Z)$ and $I_z(X,Z)$ are given respectively by
\begin{equation}
I_\rho = \int \frac{dA dT}{2\pi} \frac{X-(T-1)^2+\xi A^2}{[X-(T-1)^2+\xi A^2]^2 + (Z-2AT)^2},
\end{equation}
and
\begin{equation}
I_z = \int\frac{dA dT }{2\pi} \frac{Z-2AT}{[X-(T-1)^2+\xi A^2]^2 + (Z-2AT)^2}.
\end{equation}

In the galactic plane ($Z=0$) and assuming $\xi=1$, $I_z=0$ by symmetry and the radial component $I_\rho$ has already been evaluated in \cite{Natarajan:2007xh}:
\begin{displaymath}
I_\rho(\xi=1,X) = \left\{
\begin{array}{ll}
-\frac{1}{2} &  X<0 \\
-\frac{1}{2} + \sqrt{X} &  0<X<1 \\
\frac{1}{2} &  X>1
\end{array}
\right.
\end{displaymath}

In this paper, we aim to generalize this result to arbitrary $X$ and $Z$.  As in \cite{Natarajan:2007xh}, we will restrict our analysis to the case $\xi=1$, which is of order its expected value.  (For example, based on the IRAS map of the galactic disk, Sikivie estimated in \cite{Sikivie:2001fg} that the $n=5$ caustic ring has dimensions $p \sim 130 $ pc and $q \sim 200$ pc.  Using the relation $p/q = \sqrt{16\xi / 27}$, one then obtains $\xi \sim 0.71$).  The $\xi =1$ case has the additional advantage that the algebra is less complicated.  Generalizing to the case $\xi \ne 1$ is not difficult but rather tedious, and the method used in this paper applies straightforwardly. 

\subsection{Gravitational field for $\xi=1$}
Our goal in this section is to compute the following integrals
\begin{equation} \label{key integral}
\frac{1}{2\pi} \int dA dT \frac{X-(T-1)^2 + A^2}{(X-(T-1)^2 + A^2)^2 + (Z-2AT)^2}. 
\end{equation}
and
\begin{equation} \label{key integral2}
\frac{1}{2\pi} \int dA dT \frac{Z-2AT}{(X-(T-1)^2 + A^2)^2 + (Z-2AT)^2}. 
\end{equation}

Eq. \eqref{key integral} can be rewritten as 
\begin{eqnarray} \nonumber
\frac{1}{2\pi} \int dA dT \left[ \frac{(c_3-c_1)(c_3-c_2)}{(c_3-c_4)(c_3-c_5)(c_3-c_6)}\frac{1}{A-c_3} +  \frac{(c_4-c_1)(c_4-c_2)}{(c_4-c_3)(c_4-c_5)(c_4-c_6)}\frac{1}{A-c_4} \right. \\ \label{residue}
+ \left. \frac{(c_5-c_1)(c_5-c_2)}{(c_5-c_3)(c_5-c_4)(c_5-c_6)}\frac{1}{A-c_5}+ \frac{(c_6-c_1)(c_6-c_2)}{(c_6-c_3)(c_6-c_4)(c_6-c_5)}\frac{1}{A-c_6} \right],
\end{eqnarray}
while Eq. \eqref{key integral2} can be rewritten as
\begin{eqnarray} \nonumber
\frac{1}{2\pi} \int dA dT \left[ \frac{c_3-c_7}{(c_3-c_4)(c_3-c_5)(c_3-c_6)}\frac{1}{A-c_3} +  \frac{c_4-c_7}{(c_4-c_3)(c_4-c_5)(c_4-c_6)}\frac{1}{A-c_4} \right. \\ \label{residue2}
+ \left. \frac{c_5-c_7}{(c_5-c_3)(c_5-c_4)(c_5-c_6)}\frac{1}{A-c_5}+ \frac{c_6-c_7}{(c_6-c_3)(c_6-c_4)(c_6-c_5)}\frac{1}{A-c_6} \right],
\end{eqnarray}

where
\begin{eqnarray}
c_1 &=& -i\sqrt{X-(T-1)^2} = -c_2 
\end{eqnarray}
\begin{eqnarray}
c_3 &=& -iT + \sqrt{-T^2-(X-(T-1)^2-iZ)} 
\end{eqnarray}
\begin{eqnarray}
c_4 &=& -iT - \sqrt{-T^2-(X-(T-1)^2-iZ)} 
\end{eqnarray}
\begin{eqnarray}
c_5 &=& iT + \sqrt{-T^2-(X-(T-1)^2+iZ)} 
\end{eqnarray}
\begin{eqnarray}
c_6 &=& iT - \sqrt{-T^2-(X-(T-1)^2+iZ)} 
\end{eqnarray}
\begin{eqnarray}
c_7 &=& \frac{Z}{2T}.
\end{eqnarray}

The poles $c_3$, $c_4$, $c_5$, and $c_6$ are simple, so the A integral can be evaluated straightforwardly using residue theory.  The remaining task now is to determine the location of the poles at given $X, Z$, and $T$ to see which contributes to the A integral. 

Using identities such as $(c_3-c_5)(c_3-c_6)=2(c_3-c_1)(c_3-c_2)$ and $(c_3-c_5)(c_3-c_6)=-\frac{4iT}{\xi}(c_3-c_7)$, we have
\begin{eqnarray} \nonumber
I_\rho &=& \frac{1}{8\pi} \int dA dT \left[ \frac{1}{\sqrt{-T^2-(X-(T-1)^2 - iZ)}}\left(\frac{1}{A-c_3}-\frac{1}{A-c_4}\right) \right. \\
&& + \left. \frac{1}{\sqrt{-T^2-(X-(T-1)^2 + iZ)}} \left(\frac{1}{A-c_5}-\frac{1}{A-c_6}\right) \right],
\end{eqnarray}
and
\begin{eqnarray} \nonumber
I_z &=& -\frac{i}{8\pi} \int dA dT \left[ \frac{1}{\sqrt{-T^2-(X-(T-1)^2 - iZ)}}\left(\frac{1}{A-c_3}-\frac{1}{A-c_4}\right) \right. \\
&& - \left. \frac{1}{\sqrt{-T^2-(X-(T-1)^2 + iZ)}} \left(\frac{1}{A-c_5}-\frac{1}{A-c_6}\right) \right].
\end{eqnarray}

We generalize A to a complex variable, and use as our contour the real axis and the upper half-circle of infinite radius and counter-clockwise orientation. (Note that one could use instead the lower half-circle, and our procedure still works). The contribution from integrating over the upper-half-circle perimeter vanishes since our integrand has the form $P(A)/Q(A)$, where $P(A)$ and $Q(A)$ are polynomials, and the degree of $P(A)$ is less than that of $Q(A)$ by two.  Hence, each pole in the upper half plane would contribute $2\pi i$ times its residue to Eq. \eqref{key integral}.  Slight complication arises when the poles are on the real axis, in which case we will have to deform the contour to avoid the poles.  However, the reality of the integral ensures that another pole enters the upper half plane simultaneously as another leaves it, so that our original recipe of each pole within the contour contributing $2\pi i \times \mbox{Residue}$ remains valid. 

We now analyze the behavior of the poles as $T$ changes, given fixed $X$ and $Z$.  The imaginary parts of \eqref{key integral} $c_1$, $c_2$, $c_3$, and $c_4$ are
\begin{eqnarray}
\mbox{Im}(c_3) &=& -T + \sqrt{\frac{1}{2}\left(\sqrt{(T^2+(X-(T-1)^2))^2+ Z^2} + T^2+(X-(T-1)^2) \right)} = -\mbox{Im}(c_6)\\
\mbox{Im}(c_4) &=& -T - \sqrt{\frac{1}{2}\left(\sqrt{(T^2+(X-(T-1)^2))^2+Z^2} + T^2+(X-(T-1)^2 \right)} = -\mbox{Im}(c_5)\\ .
\end{eqnarray}
We can solve for where (as a function of $T$) each pole crosses the real axis by setting any of the above to $0$.  The resultant condition is a quartic equation of the form
\begin{equation} \label{quartic}
4T^4-8T^3+4(1-X)T^2 - Z^2 = 0.
\end{equation}
To express the four roots of this equation in a more concise form, we first define the functions
\begin{widetext}
\begin{eqnarray}
f_1(X) &=& \frac{1}{3}(2+4X) \\
f_2(X,Z) &=& 4((1-X)^2-3Z^2) \\
f_3(X,Z) &=& -128(X-1)^3 - 1728 Z^2 -1152(X-1)Z^2.
\end{eqnarray}
With these, we further introduce
\begin{equation}
f_4(X,Z) = \frac{f_3+\sqrt{-4(4f_2)^3+f_3^2}}{2}.
\end{equation}
Now the four roots are given by
\begin{eqnarray}\label{quarticroot1}
T_1&=&\frac{1}{2}\left(1-\sqrt{\frac{f_1}{2} + \frac{f_2  f_4^{-1/3}}{3} + \frac{ f_4^{1/3}}{12}} - \sqrt{f_1 -\frac{f_2 f_4^{-1/3}}{3}  - \frac{f_4^{1/3}}{12} - 2X \left(\frac{f_1}{2} + \frac{f_2  f_4^{-1/3}}{3} + \frac{ f_4^{1/3}}{12}\right)^{-1/2}} \right) \\ \label{quarticroot2}
T_2&=&\frac{1}{2}\left(1-\sqrt{\frac{f_1}{2} + \frac{f_2  f_4^{-1/3}}{3} + \frac{ f_4^{1/3}}{12}} + \sqrt{f_1 -\frac{f_2 f_4^{-1/3}}{3}  - \frac{f_4^{1/3}}{12} - 2X \left(\frac{f_1}{2} + \frac{f_2  f_4^{-1/3}}{3} + \frac{ f_4^{1/3}}{12}\right)^{-1/2}} \right) \\ \label{quarticroot3}
T_3&=&\frac{1}{2}\left(1+\sqrt{\frac{f_1}{2} + \frac{f_2  f_4^{-1/3}}{3} + \frac{ f_4^{1/3}}{12}} - \sqrt{f_1 -\frac{f_2 f_4^{-1/3}}{3}  - \frac{f_4^{1/3}}{12} + 2X \left(\frac{f_1}{2} + \frac{f_2  f_4^{-1/3}}{3} + \frac{ f_4^{1/3}}{12}\right)^{-1/2}} \right) \\ \label{quarticroot4}
T_4&=&\frac{1}{2}\left(1+\sqrt{\frac{f_1}{2} + \frac{f_2  f_4^{-1/3}}{3} + \frac{ f_4^{1/3}}{12}} + \sqrt{f_1 -\frac{f_2 f_4^{-1/3}}{3}  - \frac{f_4^{1/3}}{12} + 2X \left(\frac{f_1}{2} + \frac{f_2  f_4^{-1/3}}{3} + \frac{ f_4^{1/3}}{12}\right)^{-1/2}} \right) 
\end{eqnarray}

In the galactic plane $Z=0$, Eq. \eqref{quartic} becomes $T^4-2T^3+(1-X)T^2=0$, and can be easily factored.  The four roots are $T=0$ (twice), $T=1\pm \sqrt{X}$.  Hence, the gravitational field admits a particularly simple form in this case, as we alluded to earlier. 

When $f_4$ vanishes, Eqs. \eqref{quarticroot1} to \eqref{quarticroot4} become undefined.  This occurs when $f_2 = 0$ and $f_3<0$.  The first condition implies that $Z=\pm(1-X)/\sqrt{3}$, which means that the second condition becomes
\begin{equation}
128(X-1)^3 + 1728Z^2+1152(X-1)Z^2 > 0,
\end{equation}
which implies $X>-1/8$.  Combining these conditions, we see that roots as given by Eqs. \eqref{quarticroot1} to \eqref{quarticroot4} are ill-defined on the two lines $Z=\pm(1-X)/\sqrt{3}$, $X>-1/8$.   Note that the line corresponding to the minus sign connects the highest point ($X=-1/8, Z=3\sqrt{3}/8$) and the rightmost tip ($X=1,Z=0$) of the caustic, while the other is its vertical reflection about the $Z=0$ axis. 

To solve for the roots on these two lines, we use the facts $f_2 = 0$ and $f_3<0$ to first simplify Eq. \eqref{quartic}, from which we can solve for the four roots:
\begin{eqnarray} 
T_1&=&\frac{1}{2}\left(1-\sqrt{\frac{f_1}{2} + f_5} - \sqrt{f_1 - f_5 - 2X \left(\frac{f_1}{2}+f_5\right)^{-1/2}} \right) \\ 
T_2&=&\frac{1}{2}\left(1-\sqrt{\frac{f_1}{2} + f_5} + \sqrt{f_1 - f_5 - 2X \left(\frac{f_1}{2} + f_5\right)^{-1/2}} \right) \\ 
T_3&=&\frac{1}{2}\left(1+\sqrt{\frac{f_1}{2} + f_5} - \sqrt{f_1 - f_5 + 2X \left(\frac{f_1}{2} + f_5\right)^{-1/2}} \right) \\ 
T_4&=&\frac{1}{2}\left(1+\sqrt{\frac{f_1}{2} + f_5} + \sqrt{f_1 - f_5 + 2X \left(\frac{f_1}{2} + f_5\right)^{-1/2}} \right) 
\end{eqnarray}
where 
\begin{equation}
f_5(X) = \frac{1}{3}\left(-1-6X+15X^2-8X^3 \right)^{1/3}. 
\end{equation}
\end{widetext}

Among the four roots, only the real ones correspond to poles crossing the real-axis of the $A$-complex plane.  It turns out that qualitative differences exist between the behavior of the poles within and outside the caustic, so we have to study these two cases separately.  Within the caustic (boundary given by \eqref{boundary}), all four roots are real; outside, only two of the four roots are real.  This difference has to do with why a kink in the gravitational field appears as we cross the boundary of the caustic ring, which we saw in the case of $Z=0$ discussed earlier.

\subsubsection{Behavior of the poles within the caustic}
Within the caustic, all four roots to the quartic equation are real, implying that the poles cross the real axis a total of four times.  In particular, the pole $c_3$ would cross the real axis three times, while $c_4$ would only do so once.  Since only poles in the upper half plane contribute to the integral, we will have to partition our integral over $T$ into multiple regions accordingly, and evaluate each partition separately.  While it is nontrivial to determine precisely which of the roots of the quartic equation correspond to which pole, operationally this information is not essential, as we can always just reorder the roots and use them appropriately.  That is, we compute the four roots using Eqs. \eqref{quarticroot1} -- \eqref{quarticroot4} ($T_{1\rightarrow 4}$) and rank them in increasing order, and relabel them as $\tilde T_1 \rightarrow \tilde T_4$, where $\tilde{T_1} < \tilde{T_2} < \tilde{T_3} < \tilde{T_4}$.   The smallest root $\tilde T_1$ is the value of $T$ at which $c_4$ crosses the real axis (at fixed $X$ and $Z$), while the other three larger roots are when $c_3$ crosses the real axis.  More specifically, for $T < \tilde T_1$, both $c_3$ and $c_4$ are in the upper half plane, and they would contribute $2\pi i \times [\mbox{Residue}(c_3)+\mbox{Residue}(c_4)]$ to the integral.  As $T$ increases beyond $\tilde T_1$, $c_4$ leaves the upper half and plane, and $c_5$ enters, while $c_3$ remains.  Hence, now $c_3$ and $c_5$ would contribute $2\pi i \times [\mbox{Residue}(c_3)+\mbox{Residue}(c_5)]$ to the integral.  This holds until $T=\tilde T_2$, when $c_3$ also leaves the upper half plane and $c_6$ enters simultaneously, so now $c_5$ and $c_6$ contribute to the integral.  When $T$ gets to $\tilde T_3$, $c_3$ reenters while $c_6$ leaves.  Finally, $c_3$ leaves again for $T>\tilde T_4$.

Summarizing, at given $X$ and $Z$, the poles that are in the upper half complex A plane as a function of $T$ are given by:
\begin{itemize}
\item $T \in (-\infty, \tilde T_1) : \mbox{Im}(c_3), \mbox{Im}(c_4) > 0$ 
\item $T \in (\tilde T_1, \tilde T_2) : \mbox{Im}(c_3), \mbox{Im}(c_5) > 0$ 
\item $T \in (\tilde T_2, \tilde T_3) : \mbox{Im}(c_5), \mbox{Im}(c_6) > 0$ 
\item $T \in (\tilde T_3, \tilde T_4) : \mbox{Im}(c_3), \mbox{Im}(c_5) > 0$ 
\item $T \in (\tilde T_4, \infty) : \mbox{Im}(c_5), \mbox{Im}(c_6) > 0$ 
\end{itemize}

Note also our calculation allows for an arbitrary constant which can be fixed by taking the limit that the caustic tube shrinks to a line.  In this limit, the caustic is an infinite line at $X=Z=0$, and the density of dark matter particles is proportional to $1/\sqrt{X^2+Z^2}$.  Using Gauss' law, we can show that $\vec I = \hat r /2$, where $\hat r$ is the radial unit vector pointing away from the origin (recall that $\vec I$ and $\vec g$ point in opposite directions).  Hence, we have to add $\frac{1}{2}$ to our solution for $I^i_\rho$ and $0$ to $I^i_z$.  Alternatively, we could obtain these constants from the asymptotic behaviour of $I_\rho(\xi=1,X\rightarrow \infty)$ at $Z=0$ (and taking advantage of the rotational symmetry that emerges in this limit of vanishing cross section size).

Hence after integrating over $A$, the integral $I^i_\rho$ (inside caustic) becomes
\begin{widetext}
\begin{equation}
I^{i}_\rho = \frac{i}{4} \left(\int_{\tilde T_1}^{\tilde T_2} + \int_{\tilde T_3}^{\tilde T_4}\right) dT \left(\frac{1}{\sqrt{-2T+1-X+iZ}} + \frac{1}{\sqrt{-2T+1-X-iZ}} \right) + \frac{1}{2},
\end{equation}
which can be integrated to yield
\begin{eqnarray} \nonumber
4I^i_\rho &=& \left(\sqrt{2\tilde T_4-1+X-iZ} + \sqrt{2\tilde T_4-1+X+iZ} \right) - \left(\sqrt{2\tilde T_3-1+X-iZ} + \sqrt{2\tilde T_3-1+X+iZ} \right)  \\
&& \left(\sqrt{2\tilde T_2-1+X-iZ} + \sqrt{2\tilde T_2-1+X+iZ} \right) - \left(\sqrt{2\tilde T_1-1+X-iZ} + \sqrt{2\tilde T_1-1+X+iZ} \right) +2
\end{eqnarray}
This can be simplified further to
\begin{equation} \label{Irinside}
2I^i_\rho = Re \left(\sqrt{2\tilde T_4-1+X-iZ} - \sqrt{2\tilde T_3-1+X-iZ} + \sqrt{2\tilde T_2-1+X-iZ} - \sqrt{2\tilde T_1-1+X-iZ} +1 \right) 
\end{equation}

Meanwhile, the integral $I^i_z$ becomes
\begin{equation}
I^i_z = \frac{1}{4} \left(\int_{\tilde T_1}^{\tilde T_2} + \int_{\tilde T_3}^{\tilde T_4}\right) dT \left(\frac{1}{\sqrt{-2T+1-X+iZ}} - \frac{1}{\sqrt{-2T+1-X-iZ}} \right),
\end{equation}
which can be readily integrated to yield
\begin{eqnarray} \nonumber
4I^i_z &=& \left(-\sqrt{1-X-2\tilde T_4+iZ} + \sqrt{1-X-2\tilde T_4-iZ} \right) + \left(\sqrt{1-X-2\tilde T_3+iZ} - \sqrt{1-X-2\tilde T_3-iZ} \right)  \\
&& \left(-\sqrt{1-X-2\tilde T_2+iZ} + \sqrt{1-X-2\tilde T_2-iZ} \right) + \left(\sqrt{1-X-2\tilde T_1+iZ} - \sqrt{1-X-2\tilde T_1-iZ} \right).
\end{eqnarray}
As before, this can be simplified further to
\begin{equation}\label{Izinside}
2I^i_z = Im \left(\sqrt{2\tilde T_4-1+X-iZ} - \sqrt{2\tilde T_3-1+X-iZ} + \sqrt{2\tilde T_2-1+X-iZ}  - \sqrt{2\tilde T_1-1+X-iZ} + 1 \right).
\end{equation}
Note that $I^i_\rho$ and $I^i_z$ are respectively the real and imaginary part of the same quantity.  This is expected since the gravitational field is conservative. 
\end{widetext}

\subsubsection{Behavior of the poles outside the caustic}
Outside the caustic tube, the quartic polynomial has two complex and two real solutions.  Only the real solutions correspond to poles crossing the real axis.  If we start with $T$ very negative, only $c_3$ and $c_4$ are in the upper half plane (as Im$(c_5)=-$Im$(c_3)$ and Im$(c_6)=-$Im$(c_4)$).  As $T$ increases, first $c_4$ crosses the real axis and leaves the upper half complex plane while $c_5$ enters.  After that $c_3$ also exits while $c_6$ enters simultaneously.  We denote the two real solutions to Eqs. \eqref{quarticroot1} to \eqref{quarticroot4} as $\tilde T_1$ and $\tilde T_2$, with $\tilde T_1 < \tilde T_2$.  At given $X$ and $Z$, the poles that are in the upper half complex plane are given by 
\begin{itemize}
\item $T \in (-\infty, \tilde T_1) : \mbox{Im}(c_3), \mbox{Im}(c_4) > 0$ 
\item $T \in (\tilde T_1, \tilde T_2) : \mbox{Im}(c_3), \mbox{Im}(c_5) > 0$ 
\item $T \in (\tilde T_2, \infty) : \mbox{Im}(c_5), \mbox{Im}(c_6) > 0$.
\end{itemize}
\begin{widetext}
Using residue theory, one obtains as before that the radial component of $\vec I$ outside the caustic (as before, we add $\frac{1}{2}$ to $I^o_\rho$):
\begin{eqnarray} 
4I^o_\rho = \left(\sqrt{2\tilde T_2-1+X-iZ} + \sqrt{2\tilde T_2-1+X+iZ} \right) - \left(\sqrt{2\tilde T_1-1+X-iZ} + \sqrt{2\tilde T_1-1+X+iZ} \right) + 2.
\end{eqnarray}
Again, one can simplify this to
\begin{equation} \label{Iroutside}
2I^o_\rho = Re \left(\sqrt{2\tilde T_2-1+X-iZ} - \sqrt{2\tilde T_1-1+X-iZ} +1 \right) .
\end{equation}
The z component of $\vec I$ is 
\begin{eqnarray} 
4I^o_\rho = -\sqrt{2\tilde T_2-1+X-iZ} + \sqrt{2\tilde T_2-1+X+iZ} + \sqrt{2\tilde T_1-1+X-iZ} - \sqrt{2\tilde T_1-1+X+iZ},
\end{eqnarray}
which can be simplified to
\begin{equation} \label{Izoutside}
2I^o_z = Im\left( \sqrt{2\tilde T_2-1+X-iZ} - \sqrt{2\tilde T_1-1+X-iZ} +1 \right)
\end{equation}
\end{widetext}

\begin{figure}[ht]
\includegraphics[scale=0.7]{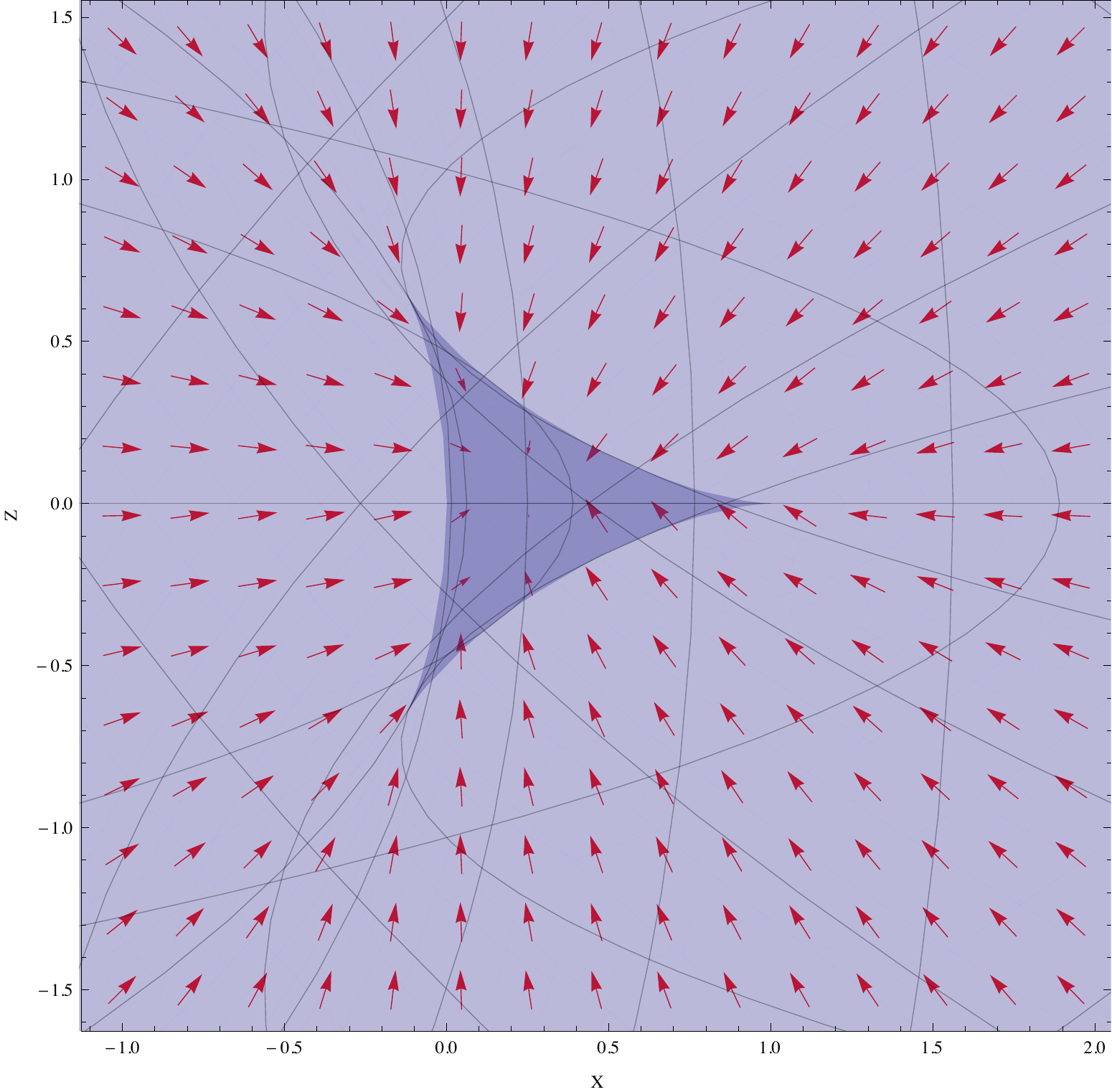}
\caption{Vector representation of $-\vec I (X,Z)$ (not $\vec g$) near a caustic ring.  A silhouette of the caustic cross section is plotted in the background by plotting parametrically Eqs. \eqref{rhoeqn} and \eqref{zeqn}. The curves in the background are contours of constant $\alpha$ and $\tau$. }
\label{gfieldvector}
\end{figure}

We plot the vector field $\vec I$ calculated using Eqs. \eqref{Irinside}, \eqref{Izinside}, \eqref{Iroutside}, \eqref{Izoutside} in Figure \ref{gfieldvector}. 

\section{Comparison with numerics}
To check the validity of our computation, we compare results obtained by our formulae with numerics.  For better convergence in our numerical calculation, we rewrite Eqs. \eqref{key integral} and \eqref{key integral2} in polar coordinates ($A=R\cos(\theta)$; $T=1+R\sin(\theta)$),
\begin{eqnarray}\label{polarintegral}
\int \frac{dR d\theta}{2\pi} \frac{ R\cos(2\theta)[X-R^2 \cos(2\theta)]}{[X-R^2\cos(2\theta)]^2 + [Z-2R\cos(\theta)(1+R\sin(\theta))]^2}, \\ \label{polarintegral2}
\int \frac{dR d\theta}{2\pi} \frac{ R\cos(2\theta)[Z-2R\cos(\theta)(1+R\sin(\theta))]}{[X-R^2\cos(2\theta)]^2 + [Z-2R\cos(\theta)(1+R\sin(\theta))]^2},
\end{eqnarray}
and first perform the integration in the angular coordinate $\theta$ in our numerical routine.  The $R$ integral is evaluated from $0$ to $100$. 

The analytical calculation is carried out using a C++ computer program, based on Eqs. \eqref{Irinside}, \eqref{Izinside}, \eqref{Iroutside}, \eqref{Izoutside}.  Due to the presence of floating-point errors, we have to introduce a threshold value in our program to decide whether a root to Eq.\eqref{quartic} is real or not: a root is considered real if its imaginary part is less than $1\times 10^{-5}$. 

\begin{figure}
\subfloat[$I_r, Z=0.1$]
{\label{fig:image_1}
\includegraphics[width=0.4\textwidth]{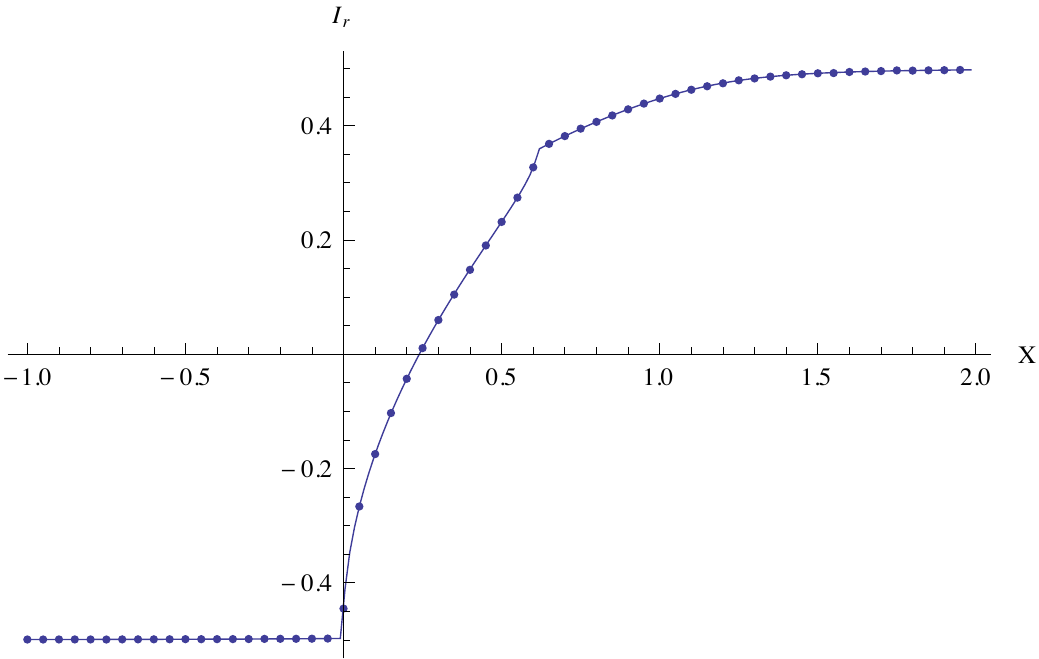}}
\subfloat[$I_z, Z=0.1$]{
\label{fig:image_2}
\includegraphics[width=0.4\textwidth]{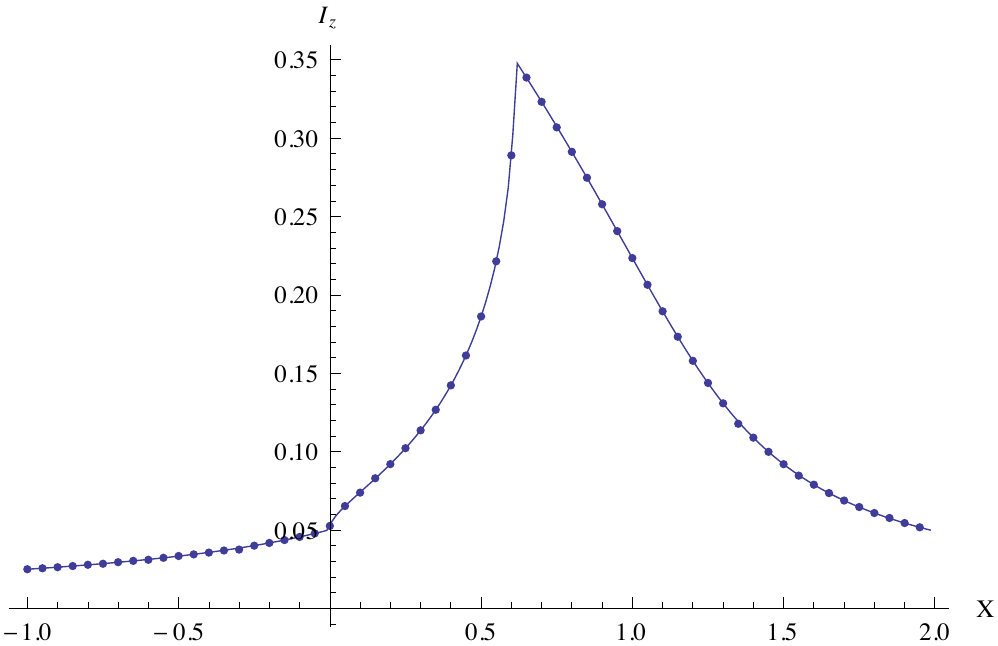}}

\subfloat[$I_r, Z=0.5$]{
\label{fig:image_3}
\includegraphics[width=0.4\textwidth]{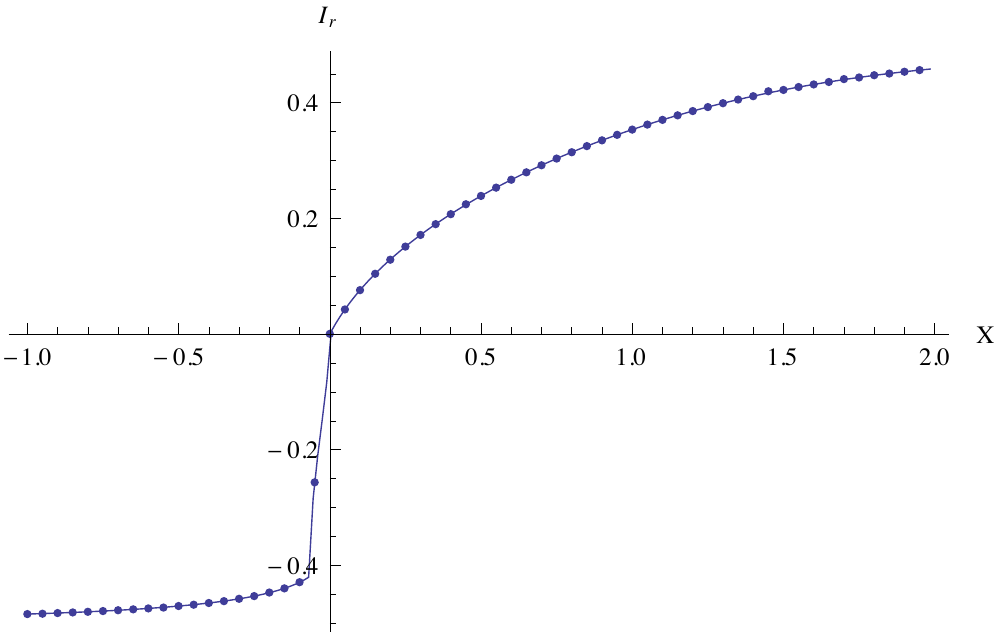}}
\subfloat[$I_z, Z=0.5$]{
\label{fig:image_4}
\includegraphics[width=0.4\textwidth]{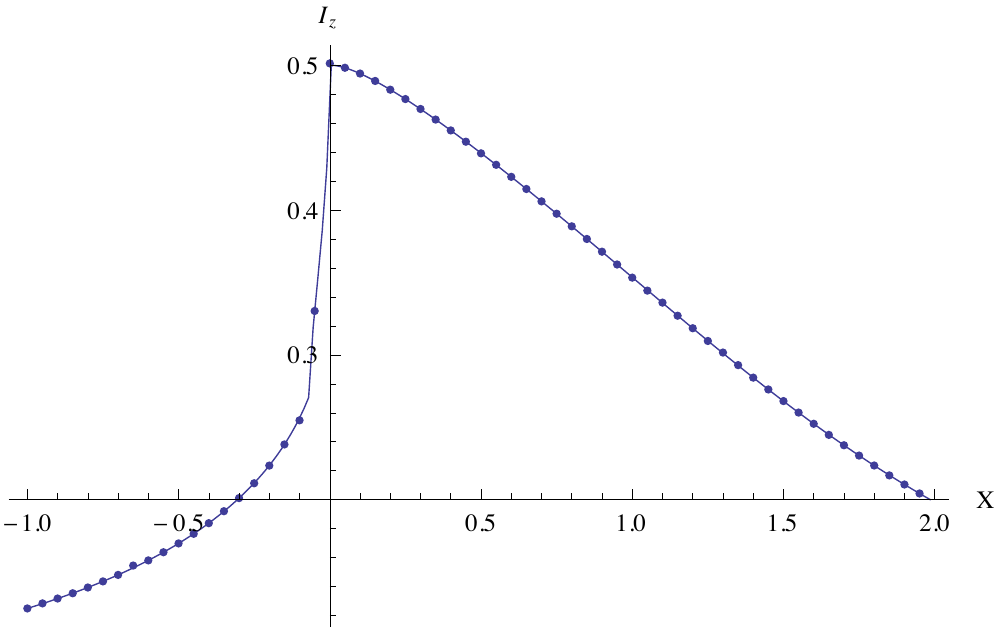}}

\subfloat[$I_r, Z=1$]{
\label{fig:image_5}
\includegraphics[width=0.4\textwidth]{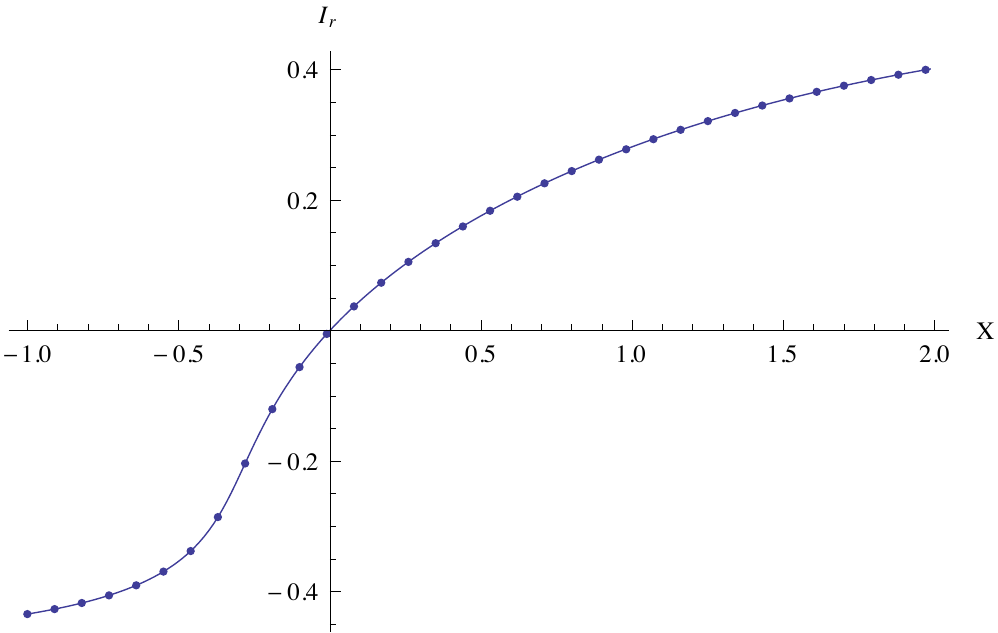}}
\subfloat[$I_z, Z=1$]{
\label{fig:image_6}
\includegraphics[width=0.4\textwidth]{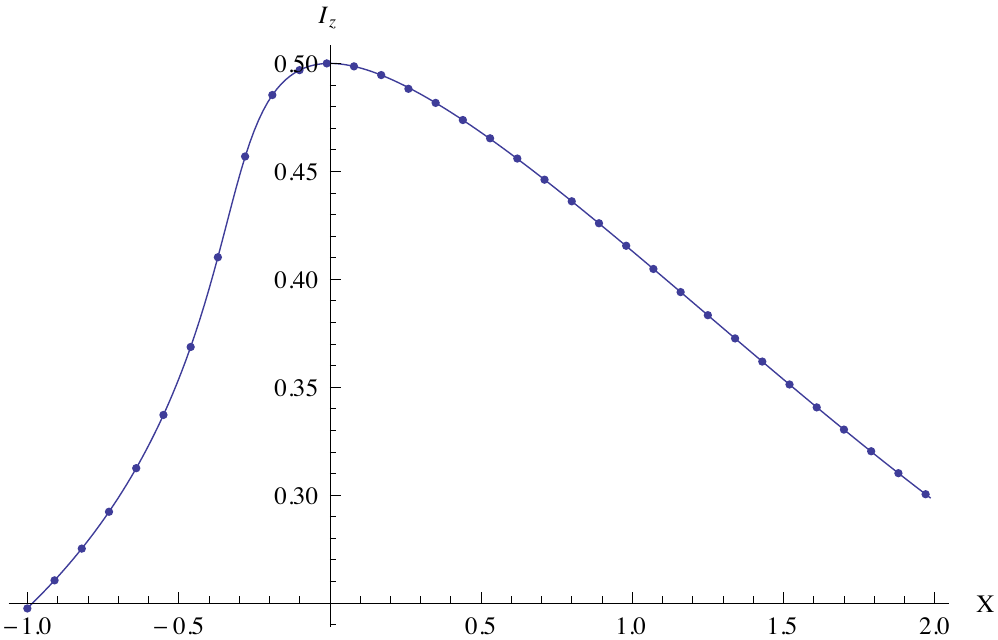}}

\caption{\label{gfield} $I_r$ and $I_z$ for $-1<X<2$ at $Z=0.1, 0.5, 1$.  The continuous curve is obtained by evaluating the analytical formulae (Eqs. \eqref{Irinside}, \eqref{Izinside}, \eqref{Iroutside}, \eqref{Izoutside}) in Sections IIB and IIC at $200$ values of $X$ distributed evenly on the interval $-1<X<2$.  The dots are calculated by numerically integrating Eqs. \eqref{polarintegral} and \eqref{polarintegral2}.  The agreement between analytics and numerics is good. }  
\end{figure}

The comparison between analytics and numerics has been checked extensively at different values of $X$ and $Z$, and is found to be in excellent agreement in every case, as is shown in Fig. \ref{gfield}, for $Z=0.1, 0.5, 1$ and $-1<X<2$.  

\section{Conclusions}
The presence of cold and collisionless dark matter particles entails the formation of dark matter caustics in galactic halos.  Evidence for inner caustic rings (as opposed to tents) exist, rendering support for the case that dark matter particles possess angular momentum and undergo net overall rotation.  This can be explained if Bose-condensed axions constitute a substantial portion of dark matter.  Due to their gravitational effects, caustic rings inevitably play an important role in the formation of structure in the early universe.  However, incorporating them into numerical simulations of structure formation is nontrivial, because the gravitational field of a caustic ring is traditionally expressed in the form of an integral, whose evaluation is computationally very expensive.  As a first step towards overcoming this difficulty, we evaluate in this paper the integral analytically in the vicinity of the ring (equivalently, taking the limit of infinite radius).  Our calculation is based primarily on the use of residue theory, in which non-analyticity in the gravitational field corresponds to poles crossing the real axis.  We find that our analytical formulae are in good agreement with numerics.  An interesting subsequent project would be to generalize our calculation to caustic rings of finite radius.

\section{Acknowledgment}
The author wishes to thank Pierre Sikivie for insightful comments, and Francisco Rojas for helpful discussions.  This work was supported in part by the U.S. Department of Energy under contract DE-FG02-97ER41029.

\end{document}